\documentclass[prd,tightenlines,showpacs,nofootinbib,eqsecnum,amsfonts,amssymb,amsmath,twocolumn,widetext]{revtex4}
\usepackage{graphicx}
\usepackage{amsmath}
\usepackage{amsfonts}   
\usepackage{amssymb}  
\usepackage{mathrsfs}
\usepackage{epsfig,bbm}
\usepackage{bm}
\usepackage{color}

\newcommand{\deu}{${\rm D}$}
\newcommand{\tro}{$^3{\rm He}$}
\newcommand{\qua}{$^4{\rm He}$}

\newcommand{\sep}{$^{7}{\rm Li}$}
\newcommand{\hui}{$^{8}$Be}

\newcommand{\carb}{$^{12}$C}
\newcommand{\hli}{$^4$He, D, $^3$He and $^{7}$Li}
\newcommand{\aaag}{$^4$He($\alpha\alpha,\gamma)^{12}$C}
\newcommand{\aabe}{$^4$He($\alpha,\gamma)^{8}$Be}
\newcommand{\beac}{$^8$Be($\alpha,\gamma)^{12}$C}
\newcommand{\hdp}{$^3$He(d,p)$^4$He}
\newcommand{\tdn}{$^3$H(d,n)$^4$He}

\newcommand{\gap}{\mathrel{ \rlap{\raise.5ex\hbox{$>$}}
                    {\lower.5ex\hbox{$\sim$}}  } }
\newcommand{\lap}{\mathrel{ \rlap{\raise.5ex\hbox{$<$}}
	            {\lower.5ex\hbox{$\sim$}}  } }

\newcommand{\etal}{{\it et al.}}

\def\iso#1#2{\mbox{${}^{#2}{\rm #1}$}}

\def\he#1{\iso{He}{#1}}
\def\li#1{\iso{Li}{#1}}
\def\be#1{\iso{Be}{#1}}

\begin{document}

\leftline{UMN--TH--3105/12, FTPI--MINN--12/19}

\title{The variation of fundamental constants and the role of $A=5$ and $A=8$ nuclei
on primordial nucleosynthesis}

\author{Alain Coc}
 \email{coc@csnsm.in2p3.fr}
 \affiliation{Centre de Spectrom\'etrie Nucl\'eaire et de
Spectrom\'etrie de Masse (CSNSM), IN2P3-CNRS and Universit\'e Paris Sud 11, UMR~8609, B\^at. 104, 91405 0rsay
Campus (France)}

\author{Pierre Descouvemont}
\email{pdesc@ulb.ac.be}
\affiliation{
Physique Nucl\'eaire Th\'eorique et Physique Math\'ematique, C.P. 229, Universit\'e Libre de Bruxelles (ULB), B-1050 Brussels, Belgium
}

\author{Keith A. Olive}
 \email{olive@physics.unm.edu}
 \affiliation{William I. Fine Theoretical Physics Institute, School of Physics and Astronomy, 
              University of Minnesota, Minneapolis, MN 55455 (USA)}

\author{Jean-Philippe Uzan}
 \email{uzan@iap.fr}
 \affiliation{Institut d'Astrophysique de Paris,
              UMR-7095 du CNRS, Universit\'e Pierre et Marie
              Curie,
              98 bis bd Arago, 75014 Paris (France)}

\author{Elisabeth Vangioni}
 \email{vangioni@iap.fr}
 \affiliation{Institut d'Astrophysique de Paris,
              UMR-7095 du CNRS, Universit\'e Pierre et Marie
              Curie,
              98 bis bd Arago, 75014 Paris (France)}

\begin{abstract}
We investigate the effect of a variation of fundamental constants on 
primordial element production in
big bang nucleosynthesis (BBN). We  focus on the effect of a possible 
change in the nucleon-nucleon interaction on nuclear reaction rates involving 
the $A=5$ ($^5$Li and $^5$He) and $A=8$ ($^8$Be) unstable nuclei and complement earlier work 
on its effect on the binding energy of deuterium.
The reaction rates for  \hdp\ and \tdn\ are dominated by the properties 
of broad analog resonances in $^5$He and $^5$Li compound nuclei
respectively.
While the triple alpha process \aaag\ is normally not effective in BBN,  
its rate is very sensitive to the position of the ``Hoyle state" 
and could in principle be drastically affected if $^8$Be were stable during BBN. 
The nuclear properties (resonance energies in  $^5$He and $^5$Li nuclei, and the binding energies of $^8$Be and D) are all computed
in a consistent way using a microscopic cluster model. The n(p,$\gamma$)d,  \hdp, \tdn\ and \aaag, reaction rates
are subsequently calculated as a function of the nucleon-nucleon interaction that can be related to the fundamental constants. 
We found that the effect of the variation of constants on the \hdp, \tdn\ and \aaag\ reaction rates is not sufficient to
induce a significant effect on BBN, even if $^8$Be was stable. 
In particular,  no significant production of carbon by the triple alpha reaction is found when compared to
standard BBN. We also update our previous analysis on the effect of a variation of constants on the n(p,$\gamma$)d reaction 
rate.
\end{abstract}

\maketitle
\today

\section{Introduction}

Constraints on the possible variation of fundamental constants are an efficient method of testing the equivalence principle~\cite{jpurmp,jpurevues}, which underpins metric theories of gravity and in particular general relativity. These constraints are derived from a wide variety of physical systems and span a large spectrum of redshifts and physical conditions, from the comparison of atomic clocks in the laboratory, the Oklo phenomena, to quasar absorption spectra up to a typical redshift of order $z\sim 2$ and big bang nucleosynthesis (BBN) at a redshift of order $z \sim10^{9}$.

Primordial nucleosynthesis is considered a major pillar of the standard cosmological model (see e.g., Refs.~\cite{books}).
Using inputs from WMAP for the baryon density~\cite{wmap7}, BBN yields excellent agreement between the theoretical predictions and  astrophysical determinations for the abundances of D and \he4~\cite{bbn2,Coc04,Ioc07,cfo5,Coc12}
despite the discrepancy between the theoretical prediction of \li7 and its determined abundance
in halo stars \cite{cfo5}. Indeed, BBN has been used extensively to constrain deviations from the standard model framework, and in particular from general relativity, see e.g., Ref.~\cite{bbnrg}.

The effects of the variation of fundamental constants on BBN predictions is difficult to model because of the intricate structure of QCD and its role in low energy nuclear reactions and because one cannot restrict the analysis to a single constant. One can, however, proceed in a two step approach: first by determining the dependencies of the light element abundances on the BBN parameters and then by relating those parameters to the fundamental constants (see Section~3.8 of Ref.~\cite{jpurmp} for an up-to-date overview). While early works have mostly focused on a single parameter such as the fine structure constant~\cite{alphaseul}, the Higgs vacuum expectation value (vev)~\cite{dixit,vvar,scherrer}  or the QCD scale~\cite{lqcdseul}, in many theories which allow for the variation of fundamental parameters, 
often, the variation of several parameters are correlated in a model dependent way \cite{wett,co}. 

The variation of a fundamental parameter such as the fine structure constant will affect the BBN analysis through the proton-to-neutron mass difference and the neutron lifetime~\cite{alphaseul} as well as the deuterium binding energy~\cite{flam1} and the binding energies of other light nuclei such as tritium, helium-3 and 4, lithium-6 and 7, and beryllium-7~\cite{other-B}. These effects can in principle be used to probe the coupled variations of several parameters~\cite{bbnmultipara,Dent-et-al,Coc07}.

Following our previous work~\cite{Coc07,Eks10,Luo11}, we allow for a variation of all fundamental constants and in order to reduce the arbitrariness, we focus on scenarios in which the variations of the different constants are correlated. In effectively all unification models of non-gravitational interactions, a variation in the fine structure constant is associated with the variation in other gauge couplings~\cite{wett,co}. Any variation of the strong gauge coupling $\alpha_s$ will induce a variation in the QCD scale, $\Lambda_{\rm QCD}$, as can be seen from the expression
\begin{equation}
 \Lambda_{\rm QCD}=\mu\left(\frac{m_{\rm c}m_{\rm b}m_{\rm t}}{\mu^3}\right)^{\frac{2}{27}}\exp\left[-
 \frac{2\pi}{9\alpha_s(\mu)}\right]
\end{equation}
valid for a renormalization scale $\mu>m_{\rm t}$. In this expression $m_{\rm c,b,t}$ are the masses of the charm, bottom and top quarks. Since the masses of the quarks are proportional to the product, $hv$, of a Yukawa coupling $h$ and Higgs vacuum expectation value $v$,  any variation of the Yukawa couplings will also induce a variation of $\Lambda_{\rm QCD}$. These variations can be related by
\begin{equation}\label{DeltaLambda}
  \frac{\Delta \Lambda}{\Lambda} = R \, \frac{\Delta \alpha}{\alpha} +
  \frac{2}{27} \left(3 \, \frac{\Delta v}{v} + \frac{\Delta h_{\rm c}}{h_{\rm c}} + \frac{\Delta h_{\rm b}}{h_{\rm b}}
  +  \frac{\Delta h_{\rm t}}{h_{\rm t}} \right) \,.
\end{equation}
The coefficient $R$ is determined by the particular grand unified theory and particle content of the theory which
control both the value of $\alpha(M_{\rm GUT}) = \alpha_s(M_{\rm GUT})$ and the low energy relation
between $\alpha$ and $\alpha_s$, leading to a considerable model dependence in its value~\cite{Dent,dine}.
Here we shall assume a typical value, $R \sim 36$~\cite{co,Langacker}. Furthermore, in theories 
in which the electroweak scale is derived by dimensional transmutation, changes in the Yukawa couplings (particularly the top Yukawa, $h_t$) lead to exponentially large changes in the Higgs vev. In such cases, the Higgs expectation value is related to the Planck mass, $M_{\rm P}$, by
\begin{equation}
 v\sim M_{\rm P}\exp\left(-\frac{2\pi c}{\alpha_{\rm t}}\right)
\end{equation}
where $c$ is a constant of order unity, and $\alpha_{\rm t}=h_{\rm t}^2/4\pi$. Thus we can write,
\begin{equation}\label{Deltav} 
   \frac{\Delta v}{v} \equiv S\,  \frac{\Delta h}{h} \,,
\end{equation}
and, as in Ref.~\cite{Coc07}, we take $S\sim240$, though there is considerable model-dependence in this value as well. For example, in
supersymmetric models, $S$ can be related to the sensitivity of the Z gauge boson mass to the top Yukawa,
and may take values anywhere from about 80 to 500~\cite{santoso}. This dependence gets translated into a variation
in all low energy particle masses~\cite{dixit}. In addition, in many string theories, all gauge and Yukawa couplings are determined by the expectation value of a dilaton. Therefore, once we allow
$\alpha$ to vary, virtually all masses and couplings are expected to vary as well, typically much more strongly
than the variation induced by the Coulomb interaction alone.

The use of coupled variations has led to significantly improved constraints in a wide range of environments 
ranging from big bang nucleosynthesis~\cite{co,bbnmultipara,Coc07,flam1,other-B,fl1,Ber10,landau,grant}, the Oklo reactor~\cite{opqccv}, meteoritic data~\cite{Dent-et-al,opqccv}, the microwave
background~\cite{landau,cmb1}, stellar evolution~\cite{Eks10} and atomic clocks~\cite{Luo11}. Concerning BBN, the effect of coupled variations has mostly focused on the binding energy, $B_D$, of deuterium~\cite{flam1,fl1,Coc07} (see also Ref.~\cite{otherbd}  for related investigations). The importance of $B_D$ is easily understood  by the fact that the equilibrium abundance of deuterium and the reaction rate $p(n,\gamma)$D both depend exponentially on $B_D$ and on the fact that deuterium is in a shallow bound state. Indeed, in Ref.~\cite{Coc07}, we found that even
a relatively small variation in the gauge or Yukawa couplings of order of a few $\times 10^{-5}$ had a significant effect on the light element abundances. In particular, using~\cite{Coc07}
\begin{equation}\label{DeltaBd3}
 \frac{\Delta B_D}{B_D} = -13 (1+S) \, \frac{\Delta h}{h}
            + 18R \, \frac{\Delta \alpha}{\alpha} \, ,
\end{equation}
a variation in the Yukawa couplings of $2 \times 10^{-5}$ induces a relative
variation in $B_D$ of about 4\%. By decreasing $B_D$, nucleosynthesis begins later at a lower temperature
ultimately suppressing the \li7 abundance.

It is well known in principle, that the mass gaps at $A = $5 and $A=8$, prevent the nucleosynthetic
chain from extending beyond \he4.  Although some \li6 and \li7 is produced,
their abundances remain far below that of the lighter elements, while B, Be, and CNO
isotopes are produced in even smaller amounts.
The presence of these gaps is caused by the instability of \he5, \li5 and \be8
with respect to particle emission: their lifetimes are as low as a few 10$^{-22}$~s for
 \he5 and \li5 and $\approx10^{-16}$~s for \be8.   
More precisely,  \he5, \li5 and \be8 are respectively unbound by 0.798, 1.69 and 0.092~MeV
with respect to neutron, proton and $\alpha$ particle 
emission\footnote{Although \li8 and ${}^8$B also contribute to 
the mass gap due to their short lifetimes (on BBN timescales), they are more deeply unbound, 
and a far greater change in the fundamental couplings would be needed to affected their stability.
We will not consider them further here.}.
Variations of constants will affect the energy levels of the unbound \he5, \li5 
and \be8 nuclei \cite{flam1,Ber10} and hence, the resonance energies whose contributions
dominate the reaction rates.
In addition, since \be8 is only slightly unbound, one can expect that 
for even a small change in the nuclear potential, it could become bound and may 
thus severely impact the results of standard BBN (SBBN), in a similar
way that a bound dineutron impacts BBN abundances~\cite{bbneffect}.
It has been suspected that stable \be8 would trigger the production of heavy elements
in BBN, in particular that there would be significant leakage of the nucleosynthetic chain 
into carbon.  Indeed, as we have seen previously~\cite{Eks10}, changes in the nuclear potential
strongly affects the triple alpha process and as a result, strongly affects the nuclear abundances
in stars.

This article investigates in detail the effect of the variation of fundamental constants on
the properties of the compound nuclei \he5, \li5, \be8 and \carb\ involved in the \tdn, \hdp\ and \aaag\
reactions, and their consequences on reaction rates and BBN abundances. In addition, we consider the 
particular case of stable \be8.

In Section~\ref{sec1b}, we briefly present the microscopic cluster model used to determine resonance
properties.
Section~\ref{sec2} focuses on \be8 and on the CNO production for the cases of both unbound and bound \be8. We compute the C/H ratio as a function of the parameter $\delta_{NN}$. Section~\ref{sec3} focuses on the \he4 production by the  \hdp\ and \tdn\ reactions. Each of these reactions contains a broad s-wave resonance at low energies, and their reaction rates may depend on the resonance energies. Section~\ref{sec4} summarizes the BBN constraints on the variation of the nuclear interaction, hence extending our previous analysis~\cite{Coc07}. Section~\ref{sec5} provides a summary and our conclusions.

\section{Outline of the nuclear model}
\label{sec1b}
We follow the formalism introduced in Ref.~\cite{Eks10} in order to model the effect of the variation 
of the nucleon-nucleon (N-N) interaction. We adopt a phenomenological description of the 
different nuclei based on a cluster model in which the wave functions are approximated by clusters of 
two or three $\alpha$ wave functions.
In a microscopic theory, the wave function of a  nucleus with nucleon number $A$, spin $J$, and total parity $\pi$ 
is a solution of a Schr{\"o}dinger equation with a Hamiltonian given by
\begin{eqnarray}\label{eq1}
H=\sum_{i=1}^A T_{i}+\sum_{i>j=1}^A V_{ij},
\end{eqnarray}
where $T_{i}$ is  the kinetic energy of nucleon $i$, and  $V_{ij}$ a
nucleon-nucleon  (N-N) interaction. In general, the potential depends on space, spin and isospin coordinates
of nucleons $i$ and $j$, and can be decomposed as
\begin{eqnarray}\label{eq2}
 V_{ij}=V^C_{ij}+V^N_{ij},
\end{eqnarray}
where $V^C$ and $V^N$ represent the Coulomb and nuclear interactions, respectively. 

Solving the Schr\"odinger equation associated with Hamiltonian (\ref{eq1}) is a difficult problem, in
particular for nuclear reactions. We use a microscopic cluster model, where the nucleons are assumed to form
"groups", called clusters. This approximation is known as the Resonating Group 
Method (RGM) \cite{WT77}, and the wave function of a two-cluster system is approximated as
\begin{equation}
\Psi^{JM\pi}={\mathcal A}\Phi_1 \Phi_2 g^{J\pi}(\rho)Y_J^M(\Omega),
\label{eq_rgm}
\end{equation} 
where $\Phi_1$ and $\Phi_2$ are the internal wave functions of the clusters (defined in the shell
model), and ${\cal A}$ is the $A$-nucleon antisymmetrizor, which accounts for the Pauli
principle. In Eq. (\ref{eq_rgm}), $g^{J\pi}(\rho)$ is the relative function, depending on the relative
coordinate $\rho$, and to be determined from the
Schr\"odinger equation. With these wave functions, any physical quantity, such as spectroscopic
properties of the nucleus, or nucleus-nucleus cross sections, can be computed.

Eq. (\ref{eq_rgm}) is written for two clusters, but the extension to three clusters is
feasible \cite{DB87}. Currently, more sophisticated microscopic models (such as the No Core Shell Model \cite{NQS09} or the variational Monte Carlo (VMC) method \cite{PW01})
are available for few-nucleon systems. These models use realistic interactions (such as Argonne
AV18 \cite{av18}) but are quite difficult to apply with nucleus-nucleus collisions. In contrast, the RGM is well adapted to the
nuclear spectroscopy and to reactions, but the use of simple cluster wave functions
(such as the $\alpha$ particle which is described by four $0s$ orbitals) makes it necessary
to adopt effective N-N interactions. We use here the Minnesota potential \cite{TLT77},
well adapted to low-mass systems. This central potential reproduces the experimental deuteron
energy. It simulates the missing tensor force by an appropriate choice of the central
interaction.
The Minnesota interaction $V_N$ is described in detail in Ref.~\cite{Eks10}.

To take into account the variation of the fundamental constants, we introduce the parameters $\delta_\alpha$ and $\delta_{NN}$ to characterize the change of the strength of the electromagnetic and nucleon-nucleon interactions respectively. This is implemented by modifying the interaction potential~(\ref{eq2}) so that
\begin{eqnarray}\label{eq2b}
 V_{ij}=(1+\delta_\alpha)V^C_{ij}+(1+\delta_{NN})V^N_{ij}.
\end{eqnarray}
Such a modification will affect $B_D$, the energy levels of $^8$Be and $^{12}$C simultaneously, as well as the resonant reactions involving $A=5$ nuclei
 such as \hdp\ and \tdn. 
If \hui\ becomes bound, one will need to calculate the two reaction rates \aabe\ and \beac.

In Ref.~\cite{Eks10}, we investigated the $^8$Be and $^{12}$C($0^+_2$) energies by scaling the
Minnesota interaction. The $\delta_{NN}$ parameter then provides a link, for the Minnesota
potential, between the deuteron energy $B_D$ and the $2\alpha$ and $3\alpha$ resonance
energies. Here we extend this idea to the \tdn\ and \hdp\ reactions. 
However both reactions are known to be dominated by a low energy $\frac32^+$ resonance 
(at $E_R^{exp}=0.048$ MeV for \he5 and $E_R^{exp}=0.21$ MeV for \li5). From simple angular-momentum
couplings it is easy to see that this resonance corresponds to an $s$ wave in the entrance
channel, and to a $d$ wave in the exit channel. Consequently the coupling between these
channels can be described by a tensor force only. As mentioned earlier, this component is
neglected in the Minnesota interaction. However, as our main interest is the variation
of the resonance energy, we used single channel $^3$He+d and $^3$H+d approximations, and
performed various calculations by modifying $\delta_{NN}$ (see Section III). This approximation
is justified by the fact that these resonances essentially have a 3+2 structure.

$\delta_{NN}$ is a phenomenological parameter that can be related to the fundamental constants 
through the dependence of the deuterium binding energy on $\delta_{NN}$. With the Minnesota
N-N interaction, we find
\begin{eqnarray}\label{BDdnn}
  {\Delta}B_D/B_D = 5.716\times\delta_{NN} \, .
\end{eqnarray}
Note that other forces may provide a slightly different dependence. However, a consistent treatment of 
$B_D$ and of resonance properties in \li5 and \he5 requires the same effective interaction,
such as the Minnesota potential.
Thus we have the possibility of relating $\delta_{NN}$ to the gauge and Yukawa couplings if one matches this prediction to a potential model via the $\sigma$ and $\omega$ meson masses~\cite{flam1,fl1,Coc07, dd} or the pion mass, as suggested in Refs.~ \cite{scherrer,pi2,pi3}.
In Ref.~\cite{Coc07}, it was concluded that
\begin{equation}
  \frac{\Delta B_D}{B_D} = 18\frac{\Delta\Lambda_{\rm QCD}}{\Lambda_{\rm QCD}} -
  17\left(\frac{\Delta v}{v}+\frac{\Delta h_s}{h_s} \right),
\end{equation}
which led to the expression in Eq. (\ref{DeltaBd3}).
Eq.~(\ref{BDdnn}) can then link any constraint on $\delta_{NN}$ to the three fundamental constants $(h_s,v,\Lambda)$.\\

\section{Primordial CNO production and \hui}\label{sec2}

CNO production in SBBN has been investigated in Ref. {\cite{Ioc07} and most recently revisited in Coc \etal~\cite{Coc12}.
The direct detection of primordial CNO isotopes seems highly unlikely with the present observational 
techniques but it is important for other applications. In particular, it may significantly affect the dynamics of population III (Pop.~III) stars since hydrogen burning in low mass Pop. III stars proceeds through the
slow pp chains until enough carbon is produced, through the triple alpha reaction,
to activate the CNO cycle. The minimum value of the initial CNO mass fraction that
would affect Pop. III stellar evolution was estimated to be 10$^{-10}$~\cite{Cas93}  or
even as low as  10$^{-12}$ for less massive stars~\cite{Eks08}.
This is only two orders of magnitude above the SBBN CNO yields obtained using current
nuclear reaction rates. 
The main difficulty in BBN calculations up to CNO is the extensive network (more than $400$
reactions) needed, including n, p, $\alpha$, but also d, t and  \tro, induced reactions 
on both stable and radioactive targets. 

CNO production (mostly \carb) in SBBN is found to be in the range CNO/H =  $(0.2-3.)\times10^{-15}$~\cite{Coc12}.
In scenarios with varying constants, this number needs to be compared with the \aaag\ production of \carb, in particular if its rate is dramatically increased by a variation of the $^8$Be ground state and Hoyle state position as considered by  Ekstr{\"o}m \etal~\cite{Eks10}. Since it is a resonant reaction, its rate is very sensitive to the nuclear interaction and we recall that a 1.5\% variation of the N-N interaction  ($-0.009<\delta_{NN}<+0.006$) would induce a change in the rate between $\approx$18 to $\approx$ 2 orders of magnitude 
for temperatures between $T$ =  0.1 to 1.0~GK (see Fig.~3 in Ref.~ \cite{Eks10}).

In this section, we investigate the effect of \hui\ on primordial CNO production. We consider two cases in which \hui\ is either unbound (\S~\ref{sec2a}) or bound (\S~\ref{sec2b}). We then derive the BBN predictions (\S~\ref{sec2c}) in each case.

\subsection{Unbound \hui}\label{sec2a}\label{s:bound}

When the N-N interaction is modified by less than 0.75\% (i.e. $\delta_{NN}<7.52\times10^{-3}$), \hui\ remains unbound w.r.t. two $\alpha$--particle emission. We can therefore take the \aaag\ rate as a function of  $\delta_{NN}$ as calculated by Ekstr{\"o}m \etal~\cite{Eks10}. We recall that, in the framework of the cluster model using the Minnesota interaction, we obtained that the energy of the \hui\ ground state with respect to the $\alpha+\alpha$ threshold was given by \cite{Eks10}

\begin{eqnarray}
-B_8 & \equiv  &E_R(^8{\rm Be}) \nonumber \\
&  = &  \left( 0.09184-12.208\times\delta_{NN}\right) \, \mathrm{MeV}
\label{eq:er8}
\end{eqnarray}
where $B_8$ is the \hui\ binding energy with respect to two alpha break-up (with this convention, $B_8<0$ for unbound \hui).
We recall that, within the same model, we obtained 
\begin{eqnarray}
E_R(^{12}{\rm C}) =  \left(0.2876-20.412\times\delta_{NN}\right) \ \mathrm{MeV}
\label{eq:er12}
\end{eqnarray}
for the dependence of the Hoyle state resonance.

\subsection{Bound \hui}\label{sec2b}\label{s:unbound}

When $\delta_{NN}\gap7.52\times10^{-3}$, \hui\ becomes bound and should be considered as a stable isotope during BBN. Hence, we have to calculate two reaction rates: \aabe\ and \beac. 
The calculation of the rate of the second reaction can be achieved using the sharp resonance formula~\cite{NACRE}  with the varying parameters of the Hoyle state from  Ekstr{\"o}m \etal~\cite{Eks10}. 
For the first reaction, \aabe,  we have performed a dedicated calculation using the potential by Buck \etal~\cite{Buc77} to obtain the astrophysical $S$-factor displayed in Figure~\ref{f:aabe1}  for values of  the \hui\ binding energy of $B_8$ = 10, 50 and 100 keV. 
The Buck potential is expressed as a Gaussian, and accurately reproduces the experimental
$\alpha+\alpha$ phase shifts up to 20 MeV. The initial $2^+$ and final $0^+$ wave functions
are computed in the potential model, assuming that the ground state is slightly bound. The cross
section is then determined from integrals involving the wave functions and the E2 operator
(see Ref. \cite{NACRE} for detail).
The broad structure corresponds to the well known 2$^+$ resonance in $\alpha-\alpha$ scattering and it is easily 
concluded from Figure~\ref{f:aabe1} that the $S$-factor remains relatively insensitive to a change in $\delta_{NN}$. 
 It was shown in Ref.~\cite{Bay85} that describing \hui\ as a bound state
($E_R < 0)$ or as a low-energy resonance ($E_R > 0$), has a small effect on the
$\alpha(\alpha,\gamma)$\hui\ capture cross section.

Figure~\ref{f:svaabe} depicts the reaction rate for  $B_8$ = 10 and 100 keV 
relative to the case with $B_8$ = 50. 
The rate depends very little on the \hui\ binding energy for $B_8 >0$ 
and the rate changes by less than $\sim$10\% for the three values of $B_8$ considered.  As a result, 
we can safely neglect the difference in the rates once $B_8 > 0$. 
The reaction rate is essentially given by the radiative capture cross section at the Gamow energy $E_0(T)$,  which is proportional to $E_{\gamma}^5$, where the photon energy is $E_{\gamma} = E_{\rm cm}+B_{\rm 8}$ and $E_{\rm cm}$ is the alpha-alpha center-of-mass energy.

\begin{figure}[htb]
 \includegraphics[width=8cm]{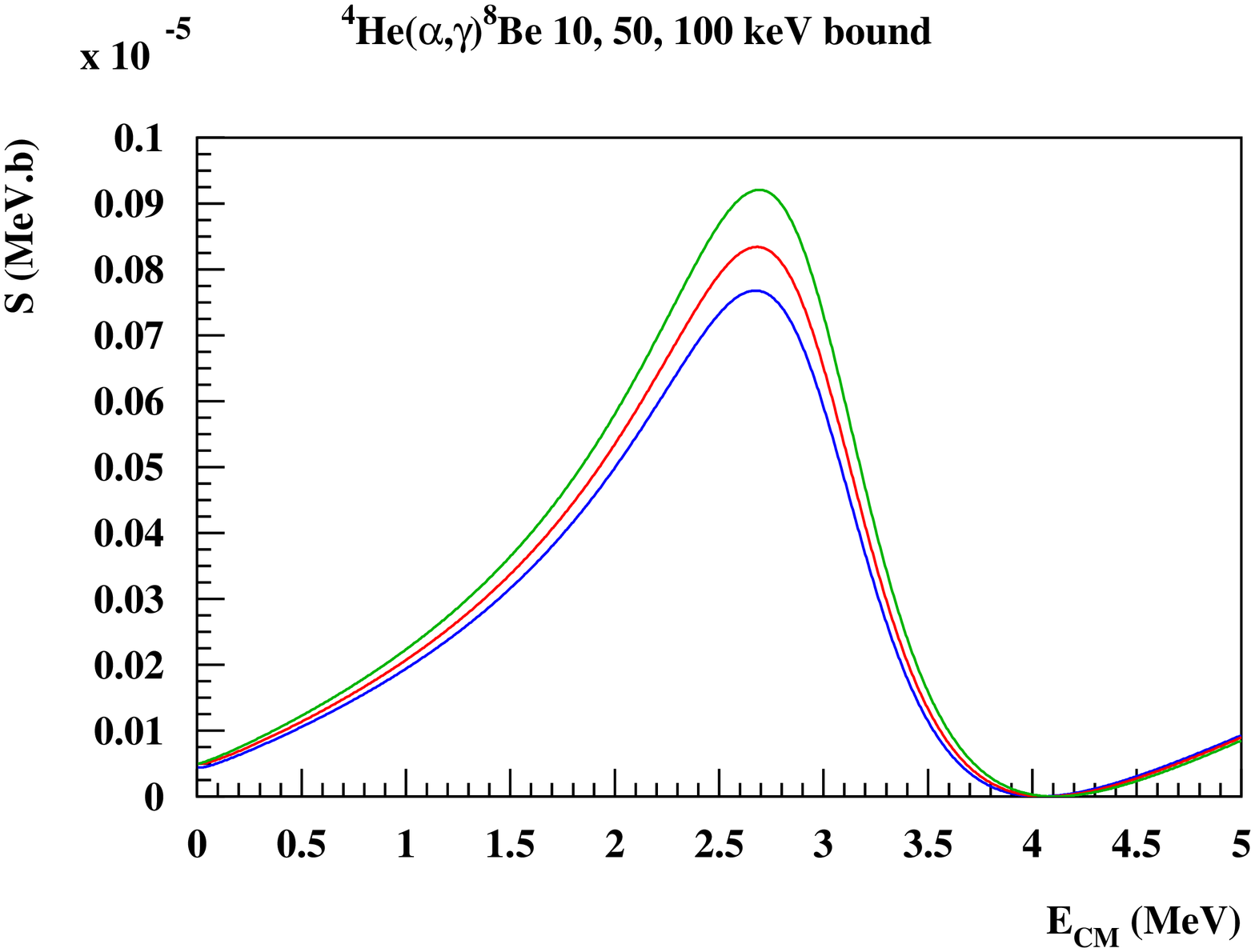}
 \caption{The astrophysical $S$-factor for the \aabe\ reaction, assuming that \hui\ is bound by 10 (blue), 50 (red) and 100 
 (green) keV, corresponding to $\delta_{NN}=0.0083, 0.0116, 0.0156$ respectively, from bottom to top.}
 \label{f:aabe1}
\end{figure}

\begin{figure}[htb]
 \includegraphics[width=8cm]{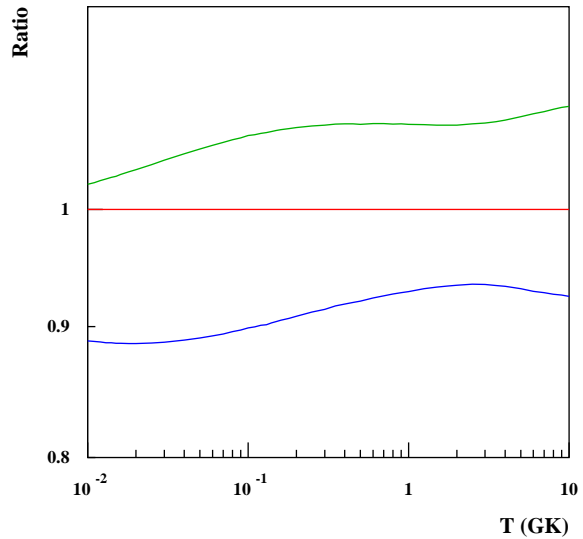}
 \caption{The relative variation of the \aabe\ reaction rate assuming that \hui\ is bound by 10 (blue), 50 (red) and (green) 100 keV, relative to the 50 keV rate.}
  \label{f:svaabe}
\end{figure}


\subsection{BBN calculations}\label{sec2c}

CNO is produced at a very low level in SBBN. The chain leading to carbon is dominated by
the following reactions: 
\begin{eqnarray}
^7{\rm Li}(\alpha,\gamma)^{11}{\rm B}  \qquad ^7{\rm Li}(n,\gamma)^8{\rm Li}(\alpha,n)^{11}{\rm B}
\label{stC1}
\end{eqnarray} 
followed by
\begin{eqnarray}
  ^{11}{\rm B}(p,\gamma)^{12}{\rm C}  \qquad   ^{11}{\rm B}(d,n)^{12}{\rm C} , \nonumber \\ 
^{11}{\rm B}(d,p)^{12}{\rm B}  \qquad ^{11}{\rm B}(n,\gamma)^{12}{\rm B}
\label{stC2}
\end{eqnarray}    
which bridge the gap between the $A\leq7$ and $A\geq12$ nuclei \cite{Ioc07,Coc12}.

To disentangle standard CNO production through the reactions listed above with the one proceeding
through the triple--alpha reaction, we reduced the network to the 15 reactions involved in $A<8$ 
nucleosynthesis plus \aaag, i.e. we turned off the other reactions, in particular those 
listed in Eqs. (\ref{stC1}) and (\ref{stC2}). As we are investigating a possible enhancement of CNO production
we only considered positive $\delta_{NN}$ values that lead to a higher triple--alpha reaction rate.

The CNO yield as a function of $\delta_{NN}$ is displayed in Figure~\ref{f:cno-dN-N}. The carbon abundance shows a maximum at $\delta_{NN}\approx0.006$, C/H$\approx10^{-21}$, which is {\em six orders of magnitude} below the carbon abundance in SBBN. 
This can be understood as follows:
Figure~\ref{f:aaag-dN-N} displays the \aaag\ rate as a function of $\delta_{NN}$ for temperatures relevant to 
 BBN, i.e., from 0.1 to 1~GK. 
 As one can see, the variation of the rate with $\delta_{NN}$ is limited at the highest temperatures
 where BBN production occurs so that
 the amplification of \carb\ production does not exceed a few orders of magnitudes. 
 Indeed, while stars can process CNO at 0.1GK over billions of years, in BBN the optimal temperature
 range for producing CNO is passed through in a matter of minutes. 
 This is not sufficient for \carb\ (CNO) nucleosynthesis in BBN.  
 Furthermore, the baryon
 density during BBN remains in the range 10$^{-5}$ to 0.1 g/cm$^3$ between 1.0 and 0.1~GK, substantially
 lower than in stars (e.g. 30 to 3000 g/cm$^3$ in Pop. III stars). This makes three-body reactions
 like \aaag\ much less efficient compared to  two-body reactions. Finally,
  in stars, \aaag\ operates 
 during the helium burning phase without significant sources of \sep, d, p and n that allow the processes
 listed in Eqs. (\ref{stC1}) and (\ref{stC2}).
 
\begin{figure}[ht]
 \includegraphics[width=8.5cm]{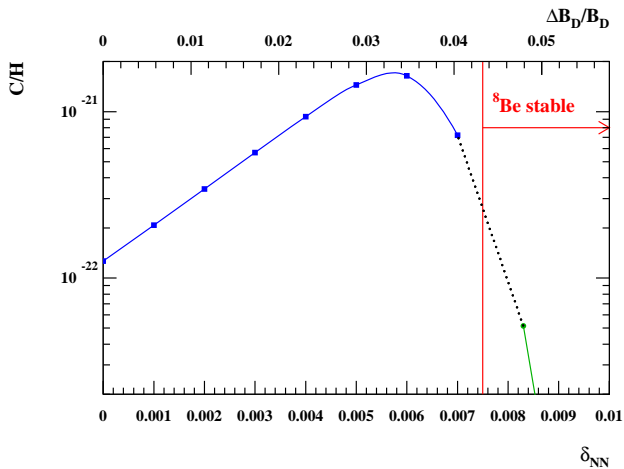}
 \caption{\carb\ production (in number of atoms relative to H) through the \aaag\ reaction (left) or via a stable \hui\ (right) as a function
 of the N-N interaction. For clarity, the rates of all other CNO producing reactions are set to zero to study these specific channels.
The dotted line just connects the results of the two types of calculations: via \aaag\ as in \cite{Eks10} and \S~\ref{sec2a} or via
$^4$He($\alpha,\gamma)^{8}$Be$^{\mathrm stable}(\alpha,\gamma)^{12}$C as in \S~\ref{sec2b}.} 
 \label{f:cno-dN-N}
\end{figure}

\begin{figure}[htb]
 \includegraphics[width=8.5cm]{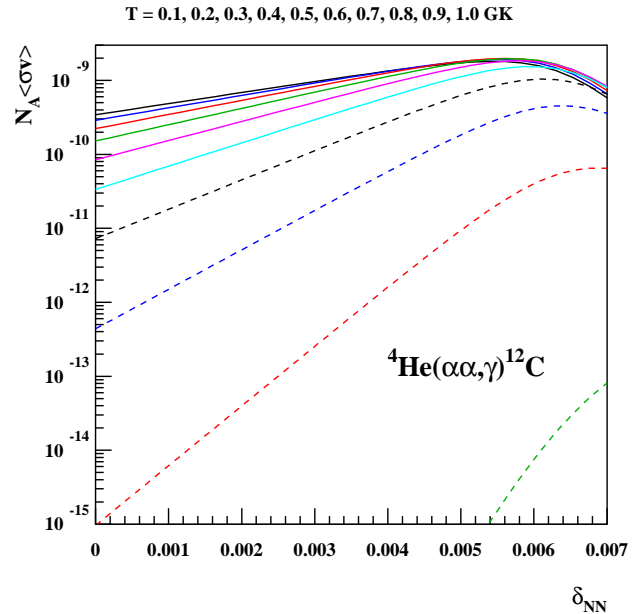}
 \caption{The \aaag\ rate as a function of $\delta_{NN}$ at constant temperature relevant for 
 BBN  from 0.1 GK (lower curve) to 1 GK (upper curve), by 0.1 GK steps.} 
 \label{f:aaag-dN-N}
\end{figure}

 The maximum of the \carb\ production as a function of $\delta_{NN}$ in Figure \ref{f:cno-dN-N} reflects the maxima 
 in the \beac\ and \aaag\ rates displayed in Figs.~\ref{f:aaag-dN-N} and \ref{f:beac-dN-N}. 
 They are due to the effect of the sharp resonances (in both the \hui\ ground state and the \carb\ ``Hoyle state"
 that  dominate the cross section).

The contribution of a sharp resonance in the $(\alpha,\gamma)$ channel is given by
\begin{eqnarray}
\langle\sigma v\rangle \propto{{ \Gamma_\alpha(E_R)\Gamma_\gamma(E_R)}\over{\Gamma(E_R)}} \exp\left(-{{E_R}\over
{\mathrm k}_BT}\right)
\label{eq:sharp}
\end{eqnarray}
where $\Gamma_\alpha$ is the entrance (alpha-)width,  $\Gamma_\gamma$ the exit (gamma-)width
and $\Gamma$ is the total width ($\Gamma= \Gamma_\alpha+\Gamma_\gamma$ if these are the only
open channels). Figure 1 in Ref.~\cite{Eks10} displays the \carb\ level scheme: the radiative width,
$\Gamma_\gamma$, is associated with the decay to the first \carb\ excited state at 4.44 MeV as
the decay to the ground state proceeds only through the much less efficient electron-positron pair
emission. The corresponding decay energy is then $E_\gamma(E_R)$ = 3.21 MeV  + ${\Delta}E_R$(\carb).    
Equation \ref{eq:sharp} shows that for a fixed $E_R$, i.e. $\delta_{NN}$, the contribution increases with temperature 
as seen in Figs.~\ref{f:aaag-dN-N} and \ref{f:beac-dN-N}.
While the radiative width $\Gamma_\gamma(E_R)\propto E_\gamma^{2\ell+1}$ is almost insensitive
 to $E_R(\delta_{NN})$,
$\Gamma_\alpha(E_R)$ is very sensitive to $E_R(\delta_{NN})$ variations because of the 
Coulomb and centrifugal barriers penetrability, $P_\ell(E)$. 
The reduced widths $\gamma_x^2$, defined by:
\begin{eqnarray}
\Gamma_x(E)=2\gamma_x^2P_\ell(E) \;\; \;\; (x\neq\gamma), 
\label{eq:reduce}
\end{eqnarray}
are corrected for these effects so that they
reflect the nuclear properties only, and are, as a good approximation, independent of $\delta_{NN}$.

Depending on whether $\Gamma_\alpha\ll\Gamma_\gamma$ or $\Gamma_\gamma\ll\Gamma_\alpha$, 
the sensitivity of $\langle\sigma v\rangle$ (Eq.~\ref{eq:sharp}) to $E_R$ or $\delta_{NN}$ variations
is very different. 
This is  due to the very different energy dependence of $\Gamma_\alpha$ and $\Gamma_\gamma$,
as discussed in detail in  \cite{New07}.
In the latter case, using Eq.~(\ref{eq:sharp}), the sensitivity of the rate to $E_R$ ($\delta_{NN}$)
variations is simply given by : 
\begin{eqnarray}
{{\partial\ln\langle\sigma v\rangle}\over{\partial\ln E_R}}=-{{E_R}\over{\mathrm k}_BT}
\end{eqnarray}
as the prefactor in Eq.~(\ref{eq:sharp}) is reduced to $\Gamma_\gamma$ which is almost constant. Since $\delta E_R$ and $\delta_{NN}$ have opposite
signs (Eqs. \ref{eq:er8} and \ref{eq:er12}), the rate {\em increases} with $\delta_{NN}$.
In the former case, the same factor is reduced to the very energy dependent $\Gamma_\alpha$ and we have: 

\begin{eqnarray}
\langle\sigma v\rangle \propto \gamma_\alpha^2\exp\left(-\sqrt{{E_G}\over{E_R}}-{{E_R}\over{\mathrm k}_BT}\right)
\label{eq:sharpa}
\end{eqnarray}
where the penetrability, $P_\ell(E)$, has been approximated by $\exp(-\sqrt{{E_G}/{E}})$ with Gamow energy,  $E_G$.
It is well known that the exponential in Eq.~(\ref{eq:sharpa}) can be well approximated (see e.g. Ref.~\cite{NACRE})
by
\begin{eqnarray}
\exp\left[-\left({{E_R-E_0}\over{\Delta E_0/2}}\right)^2\right]
\end{eqnarray}
with
\begin{eqnarray}
E_0  &=&  \left(\frac{\mu}{2}\right)^{1/3} \left( \frac{\pi e^2 Z_1Z_2 \mathrm{k}T}{\hbar}\right)^{2/3}  \nonumber \\
 &=& 0.1220\ (Z_1^2Z_2^2 A)^{1/3} \, T_9^{2/3}\  {\rm MeV}
\label{equ5}
\end{eqnarray}
and 
\begin{eqnarray}
\Delta E_0  &=&  4\ (E_0\mathrm{k}T/3)^{1/2}   \nonumber \\
                   &=& 0.2368 \ (Z_1^2 Z_2^2 A)^{1/6}\, T_9^{5/6} \ {\rm MeV}. 
\label{equ6}
\end{eqnarray}
that define the Gamow window.
Recalling that the reduced width $\gamma^2$ only reflects the nuclear structure and is assumed to 
be constant, it is straightforward to calculate the sensitivity of the rate to $E_R$ ($\delta_{NN}$)
variations:
\begin{eqnarray}
{{\partial\ln\langle\sigma v\rangle}\over{\partial\ln E_R}}=4\left({{E_0(T)-E_R(\delta_{NN})}\over{\Delta E_0(T)/2}}\right)
\end{eqnarray}
Since, for large $\delta_{NN}$ and $T$, we have $E_0>E_R$, the rate {\em decreases} with $\delta_{NN}$.

The condition $\Gamma_\alpha=\Gamma_\gamma$ that marks the boundary between these two opposite
evolutions in Eq.~(\ref{eq:sharp}) can be found in Fig.~A.1 of Ekstr{\"o}m \etal~\cite{Eks10}, at $\delta_{NN}\approx0.006$.
It also corresponds to the maximum of the \beac\ rate depicted in Fig.~\ref{f:beac-dN-N}. The more
complicated two step, three body \aaag\ reaction shows a similar dependence.

To summarize, for  $\delta_{NN}\lap0.006$, the rate decreases (increases) as a function of $E_R$ ($\delta_{NN}$)
because of the dominating exponential factor,  $\exp(-{{E_R}/{\mathrm{k}T}})$, while for $\delta_{NN}\gap0.006$, it
increases (decreases) because of the penetrability. This evolution is followed by the \carb\ production
displayed in Fig.~\ref{f:cno-dN-N}.

For $\delta_{NN}\geq0.00752$, when \hui\ is bound, \carb\ production drops to C/H $\approx$ $5\times10^{-23}$ for  $B_8$ = 10.  For larger $B_8$, the abundance drops sharply as seen in Figure \ref{f:cno-dN-N}. For $B_8$ = 50 keV, C/H $\approx$ $5\times10^{-29}$  and is
no longer in the range shown in the figure. For   $B_8$ = 100  keV, corresponding to $\delta_{NN}=0.0156$, the Hoyle state is even below threshold and the production is vanishingly small. If \hui\ is bound, reactions that normally produce  two $\alpha$-particles could form \hui\ instead. We considered the following reactions 
\begin{eqnarray}
^7{\rm Be}(n,\gamma)2\alpha \qquad ^7{\rm Li}(p,\gamma)2\alpha \nonumber \\
^7{\rm Li}(n,\gamma)^8{\rm Li}(\beta^+)2\alpha \qquad   ^7{\rm Be}(d,p)2\alpha \nonumber \\ 
^7{\rm Li}(d,n)2\alpha \qquad   ^7{\rm Be}(t,np)2\alpha  \nonumber \\
^7{\rm Be}(^3{\rm He},2p)2\alpha \qquad  ^7{\rm Be}(n,\gamma)2\alpha  \nonumber \\
^7{\rm Li}(^3{\rm He},d)2\alpha  \qquad  ^7{\rm Be}(t,d)2\alpha \nonumber \\
^7{\rm Li}(t,2n)2\alpha \qquad ^7{\rm Li}(^3{\rm He},np)2\alpha \nonumber
\end{eqnarray}
using the same rates as in Ref.~\cite{Coc12} but replacing $2\alpha$ by \hui. The only significant enhancement  comes from the $^7$Li(d,n)2$\alpha$ reaction but even in the most favorable case ($B_8$ = 10 keV), C/H reaches $\approx10^{-21}$.
This is still {\em six orders of magnitude} below the SBBN yield~\cite{Coc12} that proceeds via the reactions  listed in Eqs. (\ref{stC1}) and (\ref{stC2}).

\begin{figure}[ht]
 \includegraphics[width=8.5cm]{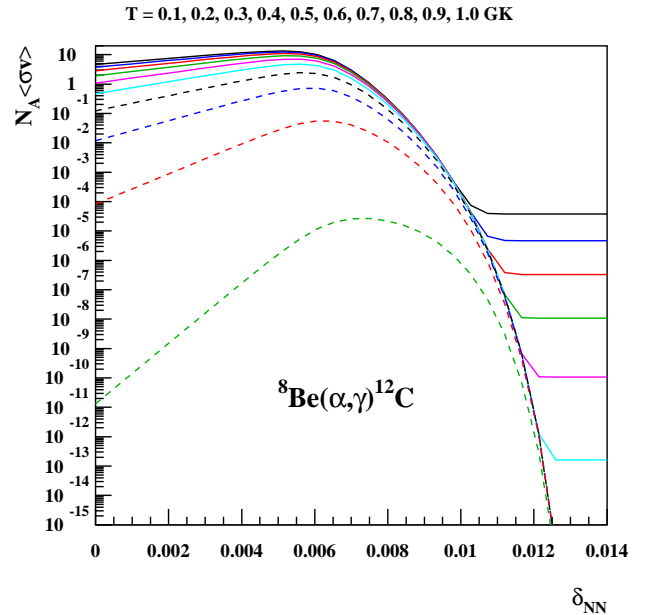}
 \caption{As in Fig.~\ref{f:aaag-dN-N} but for the \beac\  rate. 
  Above $\delta_{NN}\approx0.006$, the contribution of the Hoyle state decreases (see text),
  leaving only the contribution of  the $3^-$ resonance at $E_R$ = 2.274~MeV
 at the highest $\delta_{NN}$ values.
} 
 \label{f:beac-dN-N}
\end{figure}

Figure~\ref{f:betime} displays the  evolution of the \carb\ and \hui\  mass fractions as a function of time. They both increase with time until the \aabe\ drops with decreasing temperature. Afterwards, equilibrium between two $\alpha$--particle fusion and \hui\ photodissociation prevails as shown by the dotted lines. Indeed, the Be mass fraction is
\begin{equation}
Y_{\mathrm Be}={{Y_\alpha^2}\over2}{\rho\over{R}}  \propto  T^{3\over2}\exp\left(-B_8/{\mathrm k}_BT\right)
\end{equation}
where we have taken the reverse ratio, $R$, to be proportional to  $\exp\left(B_8/{\mathrm k}_BT\right){\times}T^{3\over2}$ and the baryon density,  $\rho\;{\propto}\;T^3$. For the highest values of $B_8$, the \hui\ mass fraction increases until, due to the expansion,
equilibrium drops out, as shown by the late time behaviour of the upper curve   ($B_8$ = 100 keV) in Figure~\ref{f:betime}.

\begin{figure}[htb]
 \includegraphics[width=10cm]{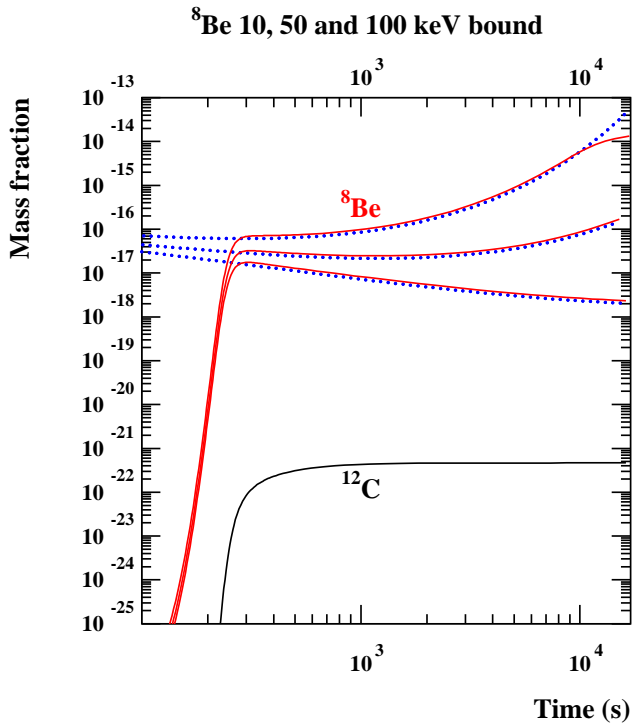}
 \caption{
 \carb\ and \hui\ mass fractions as a function of time, assuming \hui\ is bound by 100, 50 and 10 keV 
 as shown by the upper to lower curves respectively.
 (Only the \carb\ mass fraction curve,  for $B_8$ = 10  keV, is shown; others
 are far below the scale shown). The dotted lines correspond to the computation at thermal equilibrium.} 
 \label{f:betime}
\end{figure}

In conclusion, if one keeps only the first 15 nuclear reactions while adding the possibility of primordial carbon production through $3\alpha$-process, one gets a typical abundance of only order C/H$\sim10^{-21}$ which remains six orders of magnitude smaller than the SBBN carbon production, which is of order C/H$\sim10^{-15}$. This explicitly shows that CNO production in BBN cannot be increased through either an increase of the $3\alpha$-reaction rate or even through the stabilization of \hui.

\section{Reactions involving $A=5$ nuclei}
\label{sec3}
\subsection{Introduction}

In contrast to the CNO elements, the \hli\ isotopes are produced in observable quantities in BBN and can thus constrain $\delta_{NN}$. 
It would be desirable to have the dependence of each of the main SBBN reactions to the N-N interaction
or other fundamental quantities. 
This was achieved in Ref.~\cite{Coc07} for the first two BBN reactions: the n$\leftrightarrow$p weak interaction and the p(n,$\gamma$)d 
bottleneck. 

Here, we propose to extend this analysis to the \tdn\ and \hdp\ reactions that proceed through the $A=5$ compound nuclei $^5$He and
$^5$Li.
In these two reactions, the rates are dominated by the contribution of a $\frac32^+$ resonance  whose 
properties can be calculated within the same microscopic model that we used for \aaag, but with  
$^3$H+d and $^3$He+d cluster structures.   
Unlike the case for $^8$Be, the lifetime of the $^5$He and
$^5$Li states is extremely short (the width of the $^8$Be ground state is 6 eV, whereas the widths
of the $\frac32^+$ resonances in $^5$He and $^5$Li are of the order of 1 MeV).  
Therefore the issue of producing $A=5$ bound states, or even shifting their ground state energy down to the Gamow window, 
is not relevant.
Even a two step process, like the triple alpha reaction, where $^5$He or $^5$Li in thermal equilibrium would capture
a subsequent nucleon to form $^6$Li is completely negligible because they are unbound by $\sim$1~MeV compared
with the 92~keV of \hui. Hence no significant equilibrium abundance of $A=5$ nuclei can be reached.

The analysis of the effect of  $\delta_{NN}$ on the other rates requires additional effort, but should be smaller because
important reactions like $^3$He($\alpha,\gamma)^7$Be do not display resonant behaviour and the $S$--factor does not change much with energy (see e.g. Fig. 1h in  Ref. \cite{Des04}).
In order to proceed, we used the following procedure
\begin{itemize}
\item
We performed single-channel $^3$H+d and $^3$He+d calculations of the resonance energy as a function of $\delta_{NN}$.
\item 
We analyzed the experimental cross sections within an $R$-matrix approach.
\item
With the energy dependence obtained from the first step, we computed the reaction rate for various $\delta_{NN}$ or $B_D$
values.
\end{itemize}

\subsection{Resonance energy in $^5$He and $^5$Li}

In principle an RGM calculation of the  \tdn\ and \hdp\ cross sections could be performed as
it was done for the $2\alpha$ and $3\alpha$ systems. The main difference is that each reaction involves
two channels, and their cross section is dominated by a low-energy $\frac32^+$ resonance. As mentioned before,
a central N-N interaction, such as the Minnesota potential, does not provide a coupling between the $^3$H+d and
$\alpha$+n (or mirror) channels, since the tensor force is missing. Consequently, the neutron width is strictly zero, and
obtaining a realistic $S$--factor with the Minnesota interaction is not possible. We therefore use the RGM to estimate 
the sensitivity of the resonance energy as a function of $\delta_{NN}$.

Using the parameterization (\ref{eq2b}) for the nucleon-nucleon interaction, we modify the resonance energy.
Both the excitation energies of the $\frac32^+$ resonance  and of the thresholds vary. We find
\begin{equation}
\label{he5}
{\Delta}E_R=-0.327\times\delta_{NN}
\end{equation} 
for \tdn\ and 
\begin{equation}
\label{li5}
{\Delta}E_R=-0.453\times\delta_{NN}
\end{equation} 
for \hdp\ (units are MeV).  These energy dependences are much weaker ($\sim$ 20--30~keV for $|\delta_{NN}|\leq0.03$) than
for $^8$Be and $^{12}$C (see Eqs.~(\ref{eq:er8}) and (\ref{eq:er12})). This is expected for broad resonances
which are weakly sensitive to the nuclear interaction \cite{WT77}.

In contrast, Berengut {\em et al.}~\cite{Ber10} find a stronger energy dependence. These authors perform
VMC calculations with realistic N-N interactions, which provide better d and $^3$H/$^3$He wave functions.
However the VMC approach is not well adapted to broad resonances, such as those observed in $^5$He and
$^5$Li. Our cluster model, although using a simpler N-N interaction, is more suited to unbound states
since the asymptotic Coulomb behaviour of scattering states is exactly taken into account.

\subsection{$R$-matrix fits of the \tdn\ and \hdp\ $S$-factors}

To be consistent with our previous work, we want to reproduce, for $\delta_{NN}$=0, the  experimental $S$--factors obtained (see references in Ref.~\cite{Des04})
by a full $R$--matrix analysis, but for convenience we restrict ourselves here to the single pole $R$--matrix approximation which 
will be shown to be sufficient. 
For the \tdn\ reaction, we  use the parameterization of Barker~\cite{Bar97}, which reproduces the resonance corresponding to the
corresponding to the $\frac{3}{2}^+$ state at 16.84 MeV. 
Because the widths are energy dependent, the maximum of the cross section at $E_R^{exp}$ = 0.048 MeV, 
differs from the minimum of the denominator that appears in the  single pole $R$--matrix prescription:
\begin{equation}
\sigma(E)\propto\frac{(\hbar c)^2}{\mu E}\frac{\Gamma_{\rm in}(E)\Gamma_{\rm out}(E)}{(E_R^*+\Delta E_R^*-E)^2 + \Gamma^{2}(E)/4}
\label{eq:sigma}
\end{equation}
where $\mu$ the reduced mass, $\Gamma=\Gamma_{\rm in}+\Gamma_{\rm out}$ and we have dropped numerical factors.
The astrophysical $S$--factor displayed in Figure \ref{f:tdn} is just $\sigma(E)E\exp\sqrt{(E_G/E)}$
where the exponential factor approximates the $\Gamma_{in}$ energy dependence due to the penetrability.
The widths, $\Gamma_x$, are related to the reduced widths $\gamma_x^{2}$ by  Eq.~(\ref{eq:reduce}), and $\Delta E_R^*(E_R)$
is given by Eqs. (5.5) in Descouvemont \& Baye \cite{Des10}.
Note that $\Delta E_R^*$ is the {\em shift factor} of $R$--matrix theory, not related to the variation of constants, and hence is 
different from the $\Delta E_R(\delta_{NN}$), that we will discuss below.
Using  $E_R^*=0.091$~MeV, the reduced entrance width $\gamma_{\rm d}^{2}=2.93$~MeV and  the
reduced exit width $\gamma_{\rm n}^{2}=0.0794$~MeV provided by Barker ~\cite{Bar97}, our
results (not a fit) are in perfect agreement with the $R$--matrix fit \cite{Des04} of the experimental data shown in Figure~\ref{f:tdn} that we used in previous 
work \cite{Coc04,Coc07,CV10,Coc12}. 

Barker was also able to reproduce the existing experimental data for the \hdp\ cross section using the same parameters for the $\frac{3}{2}^+$ state at 16.87 MeV with $E_R^{exp}$ = 0.21 MeV  as seen in Figure \ref{f:hdp}. However, when including modern data as in Descouvemont \etal~\cite{Des04} the agreement was found to be poor. Accordingly, we have performed a fit of the $S$--factor provided by Ref.~\cite{Des04} that gave $E_R^*=0.35779$~MeV, with $\gamma_{\rm he-3}^{2}=1.0085$~MeV and  $\gamma_{\rm p}^{2}=0.025425$~MeV. Hence, our new calculations of the \hdp\ and \tdn\ rates coincide with those we have been using
in previous papers, when the constants do not vary, showing that the single pole approximation is sufficient.

\begin{figure}[htb]
 \includegraphics[width=9cm]{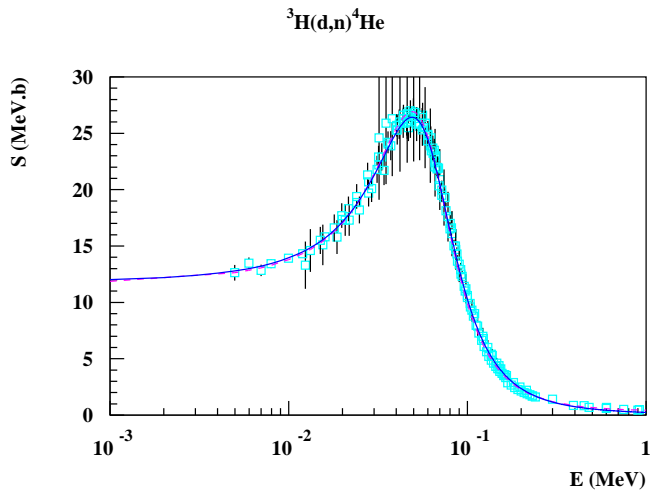}
 \caption{$S$-factor for the \tdn\ reaction from Ref.~\cite{Des04} (dashed) and our calculation using the parameters from Barker \cite{Bar97} (solid).
See Ref.~\cite{Des04} for the references to experimental data. 
 } \label{f:tdn}
\end{figure}

\begin{figure}[htb]
 \includegraphics[width=9cm]{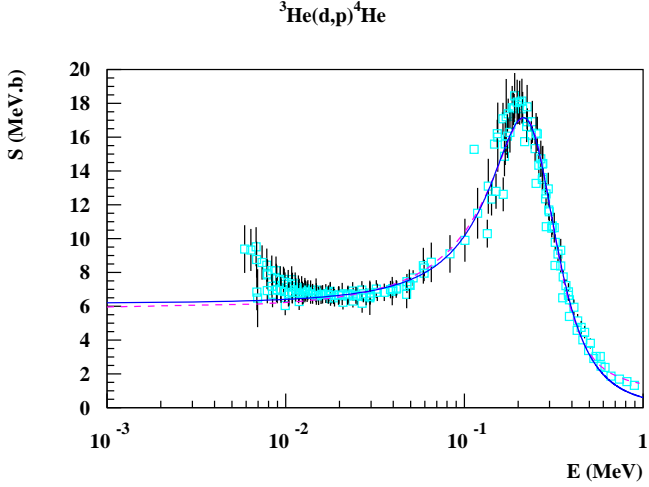}
 \caption{As in Fig.~\ref{f:tdn} but for \hdp\ and our calculation with parameters fit to ref.~\cite{Des04} (dashed). Deviations with experimental data 
 at very low energy are due to screening.  
 } \label{f:hdp}
\end{figure}

\subsection{Reaction rates}
The reaction rates are shown in Figure~\ref{f:rsv}. As expected from  Eqs. (\ref{he5}) and (\ref{li5}) they
are only slightly affected by variations of $\delta_{NN}$ (less than 5\%).  From the sensitivity study of Ref.~\cite{CV10}, we deduce that \tdn\ rate variations have no effect on BBN while  \hdp\ rate variations induce only very small ($\leq$4\%) changes in the \sep\ and $^3$He abundances.  Because the change is so small, we can make a linear approximation to the sensitivity as shown in Table~\ref{t:sensitivity} displaying $(\delta{Y}/Y)/\delta_{NN}$ values for both reactions.

\begin{figure}[htb]
\includegraphics[width=9cm]{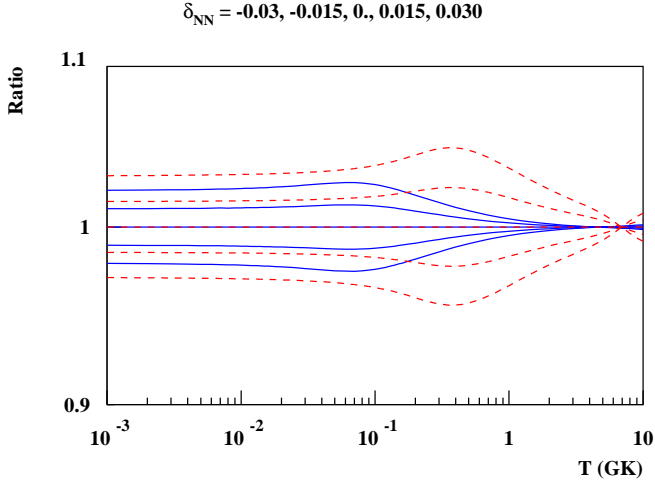}
 \caption{Relative variation of the \tdn\ (solid line) and \hdp\ (dashed line) rates for $\delta_{NN}$ = -0.030, -0.015, 0.015 and 0.030.} 
 \label{f:rsv}
\end{figure}

\begin{table}[h]
\caption{\label{t:sensitivity}Abundance sensitivity, ${\partial\log}Y/\delta_{NN}$, to a variation of the N--N interaction at WMAP baryon density. Blank entries correspond to negligible values.} 
\begin{center}
\begin{tabular}{ccccc}
\hline
Reaction  & $Y_p$   &  \deu/H &   \tro/H  &  \sep/H \\
\hline
\tdn\ & &&& -0.015 \\
\hdp\  & & -0.027 & -1.14 & -1.10 \\
\hline
 \end{tabular}\\
\end{center}
\end{table}

Our results are significantly different from those of  Berengut  \etal~\cite{Ber10}: first, we use a more elaborate parameterization of the cross--section,
second, our calculation of the resonance energy shift is less sensitive to the variation of constants. 

While, for the nominal values of the constants, both parameterizations reproduce the experimental data very well, their comparison
(our Eq.~(\ref{eq:sigma}), and Eq.~(9) Ref.\cite{Ber10}) shows important differences. Only the partial width energy dependence in the
entrance channel ($\Gamma_{\mathrm in}=\Gamma_d$) is considered in Ref.~\cite{Ber10}, neglecting the outgoing and total energy dependences. 
To be consistent, the latter must indeed depend on energy at least through $\Gamma_{\mathrm in}$. 
The partial width  ($\Gamma_{\mathrm out}=\Gamma_p$ or  $\Gamma_n$) energy  dependence in the exit channel  is small 
(large outgoing energy) and can be neglected. But, in addition,  as shown in Eq.~(\ref{eq:sigma}), the total width  
 ($\Gamma=\Gamma_d+\Gamma_p$ or  $\Gamma_d+\Gamma_n$) energy dependence (essentially from $\Gamma_d$) must be included.    
We also include in our calculations the energy shift  $\Delta E_R^*(E_R)$.
As can be seen in Figure~\ref{f:hdp1}, the variation of the S-factor with ${\Delta}E_R$  (now the supplementary
shift due to variation of constants i.e. $\neq\Delta E_R^*(E_R)$) 
is very different from ours when following Berengut \etal~\cite{Ber10} prescription. (To emphasize the different
behaviour, we have explored, in that figure a wider range of   ${\Delta}E_R$ values $\pm$ 135~keV.)

\begin{figure}[htb]
 \includegraphics[width=9cm]{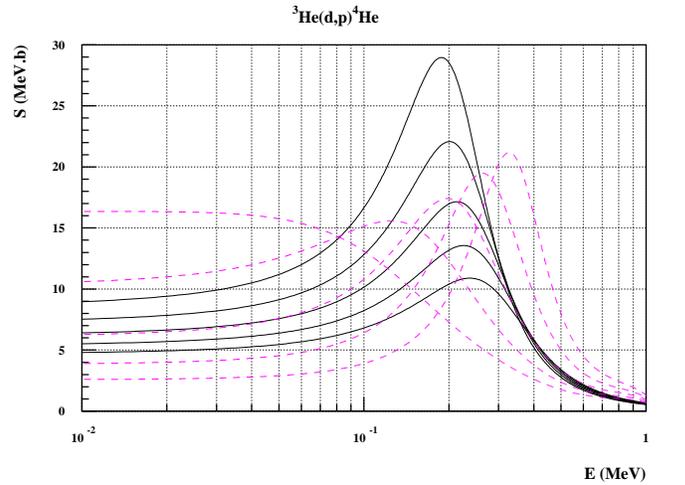}
 \caption{$S$-factor for the \hdp\ reaction and $\delta_{NN}$ = $-0.30$, $-0.15$, 0., 0.15 and 0.30 according to our calculation (solid black) or following the parameterization of Ref.~\cite{Ber10} with the same energy shifts (dashed pink). To emphasize the effect we use larger variations, by a factor of 10,  than in Fig.~\ref{f:rsv}.} 
 \label{f:hdp1}
\end{figure}

These large differences reflect the differences in the parameterization of the cross section. 
This shows that 
the expression of the cross section provided by Cyburt \cite{Cyb04} and used by Berengut  {\em et al.}~\cite{Ber10} is
excellent for a parameterization of the experimental data, it is by no means adapted to a study of the influence
of nuclear parameters on the cross--section.

The calculation of ${\Delta}E_R$ as a function of $\delta_{NN}$ is obtained by the {\em difference} between the energy
of the ${3\over2}^+$ states and of the two clusters emission thresholds in the entrance channel, both depending on the
N--N-interaction.  Berengut {\em et al.}~\cite{Ber10} assume that these levels follow the dependence of the $^5$Li and $^5$He ground 
states but these ${3\over2}^+$ state have indeed a \tro$\otimes$d or t$\otimes$d structure so that our model is more appropriate (see Ref.~\cite{bogdanova91}).

Furthermore, we note that significant changes in element abundance's occur when the variation in
the quark mass, $\delta m_q/m_q$ (their parameter of variation), is of order 1 \%. We recall that 
$\delta m_q/m_q$ can be related to a change in a Yukawa coupling through $(1+S) \delta h/h$.  
Using Eqs.~  (\ref{DeltaBd3}) and (\ref{BDdnn}) we can further relate $\delta m_q/m_q$ to $\delta_{NN}$ (ignoring the contribution from $R \delta \alpha/\alpha$) as
\begin{equation}
\delta m_q/m_q \approx -0.4 \delta_{NN}.
\end{equation}
Thus for $\delta_{NN} < 0.01$, we are essentially restricting $\delta m_q/m_q < 0.004$
and would expect a significantly smaller effect on the abundances than seen in Ref.
~\cite{Ber10}.

\section{BBN constraints}\label{sec4}

\subsection{Observational constraints}

Deuterium, a very fragile isotope, is systematically destroyed after BBN.
Its most primitive abundance is determined from the observation of
clouds at high redshift, along the line of sight of distant quasars. Very few observations of these
cosmological clouds  are available \citep{bt98a} and a weighted mean of this data yields a D/H abundance of 
\begin{equation}
\hbox{D/H} = (3.02 \pm 0.23) \times 10^{-5}. 
\end{equation}

In contrast, after BBN, \qua\ is produced by stars. Its primordial abundance is deduced from observations
in H{\sc ii} (ionized hydrogen) regions of low metallicity blue compact galaxies. 
The primordial \qua\ abundance $Y_p$ (mass fraction) is given by the extrapolation to zero
metallicity but is affected by systematic uncertainties such as plasma temperature
and underlying stellar absorption \cite{osk}. 
Using the Markov Chain-Monte Carlo methods 
described in Aver {\em et al.}~\cite{aos2} and data
compiled in Izotov \etal~\cite{its}, Aver {\em et al.}~\cite{aos3} found
\begin{equation}
\hbox{Y}_p = 0.2534 \pm 0.0083.
\end{equation} 
Given the uncertainty, this value is consistent with the BBN prediction.

 \tro\, on the other hand, is both produced and destroyed in stars so that the evolution of its
abundance as a function is subject to large uncertainties
 and has only been observed in our Galaxy \citep{Ban02},
\begin{equation}
\hbox{\tro\ /H}= (1.1 \pm 0.2) \times 10^{-5}.
\end{equation}
Consequently, the baryometric status of \tro\ is not firmly established \citep{vang03}.

The primordial lithium abundance is deduced from observations of low metallicity stars in the
halo of our Galaxy where the lithium abundance is almost independent of metallicity, displaying
a plateau, the so-called Spite plateau \citep{Spite82}.
This interpretation assumes that lithium has not been depleted on the surface of these stars, so
that the presently observed abundance is presumably primordial. The small
scatter of values around the ÒSpite plateauÓ is an indication that depletion may not have been
very effective.

Astronomical observations of these metal poor halo stars \citep{Ryanetal2000}
 have led to a relative primordial abundance of
\begin{equation}
\hbox{Li/H} = (1.23^{+0.34}_{-0.16}) \times 10^{-10}.
\end{equation}
A more recent analysis by Sbordone \etal~\cite{Sbo10} gives
\begin{equation}   
\hbox{Li/H} =  (1.58 \pm 0.31) \times 10^{-10}.
\end{equation}
For a recent review of the latest 
Li observations and their different astrophysical aspects, see Ref.~\cite{Spite10}.

\subsection{Revisiting SBBN with variable couplings}

The results of the former sections can be implemented in a BBN code in order to compute the primordial abundances of the light elements as a function of $\delta_{NN}$. We can rephrase our analysis of Ref.~\cite{Coc07} in terms of $\delta_{NN}$ by using Eqs.~(\ref{DeltaBd3}) and~(\ref{BDdnn}) which yields the constraints
\begin{equation}
 -0.7\%<\delta_{NN}<+0.5\%,
\end{equation}
assuming $R = 36$ and $S=240$. 
To test the importance of the variations in the $A=5$ rates, we first include only those variations in  \hdp\ and \tdn. 
Figure~\ref{f:a5} compares the BBN predictions for different values of $\delta_{NN}$ up to 30\%. We emphasize that a 30\% variation in $\delta_{NN}$ is unrealistic since it corresponds to a 175\% variation on $B_D$. As one can see, the curves for \he4 and  D/H are nearly horizontal, and the effect on \li7 is
insufficient to solve the lithium problem.

Next, we allow all rates depending on $\delta_{NN}$ to vary. Thus, 
Figure~\ref{f:ha240} updates Figure~4 of Ref.~\cite{Coc07} assuming again $S=240$ and $R=36$.
 In this case, we see that the \deu\ and \qua\ abundances 
can provide constraints on the variations of $h$, compatible with zero, while the \sep\ abundance can only be reconciled with observations
for $\delta h/h\approx+3\times 10^{-5}$. 
We now find 
\begin{equation}
-2 \times 10^{-6}< \frac{\delta h}{h}< 8 \times 10^{-6}.
\end{equation}
Recalling that with $S=240$ and $R=36$, one has
\begin{eqnarray}
 \delta_{NN} \sim -321 \frac{\delta h}{h},
 \label{eq:h2dN-N}
\end{eqnarray}
which gives
\begin{equation}
-0.0025 <\delta_{NN}< 0.0006.
\end{equation}
Thus, we see that variations in $\delta_{NN}$ as large as that needed to reconcile \li7
induce an excess of D/H and a deficit of \he4 (even when the large uncertainties in $Y_p$ are
taken into account).

\begin{figure}[htb]
 \includegraphics[width=7cm]{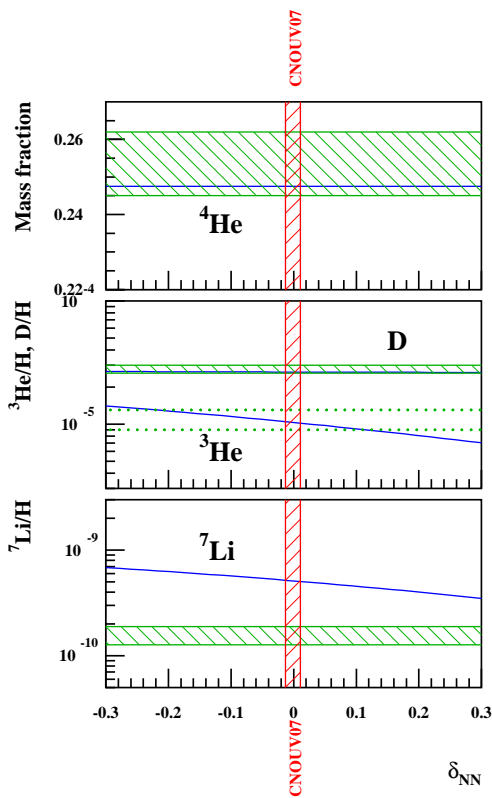}
 \caption{Effect of the variation of the N-N interaction induced solely by the modification of the nuclear rates of  \hdp\ and \tdn\  on the primordial abundances of the light element compared to the constraints obtained in Ref.~\cite{Coc07}.} 
 \label{f:a5}
\end{figure}

\begin{figure}[htb]
 \includegraphics[width=7cm]{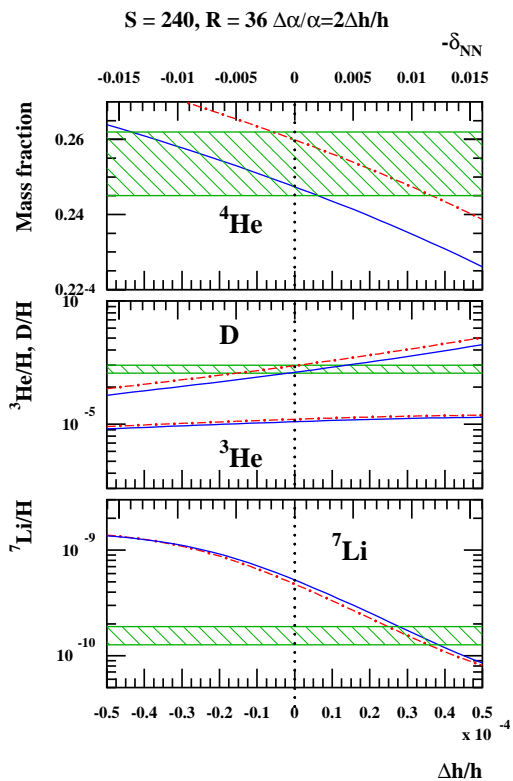}
 \caption{Update Figure~4 of Ref.~\cite{Coc07} assuming $S=240$ and $R=36$ (solid blue line),
 using new rates for $^{3}$He($\alpha,\gamma)^{7}$Li  \cite{Cyb08a} and $^{1}$H(n,$\gamma)$D \cite{And06} and 
 the $\Omega_b$ value from WMAP7 \cite{wmap7}. The top axis is $-\delta_{NN}$ from Eq.~(\ref{eq:h2dN-N}) (mind the sign) and
 the dashed red line assumes $N_\nu=4$. }
\label{f:ha240}
\end{figure}

It is interesting to note that the latter problem can be reconciled by adding a relativistic degree of freedom.
Indeed, it has recently been pointed out that extra relativistic degrees of freedom may be required from the analysis of CMB data~\cite{nu1} and BBN data~\cite{nu2}.
As a consequence, we also show in Figure~\ref{f:ha240}, the resulting abundances
when number of relativistic species is equivalent to four neutrino families.
As one can see, even at large $\delta_{NN}$, the helium abundance is concordant with observations.
Interestingly, although the constraint from \he4 on $\delta h/h$ for positive values is relaxed,
the constraint from D/H is tightened and becomes more severe for $N_\nu = 4$. 
Therefore, to reconcile
D and \li7 abundances we must
rely on subsequent destruction of D/H to match the quasar absorption system data.
In fact, high D/H is a general consequence of lowering \li7 from SBBN values \cite{opvs}.
For $\delta h/h < 0$, while the constraint from D/H is relaxed, the constraint from
\he4 is strengthened and a similar limit is obtained.

\section{Discussion}\label{sec5}

This article investigated the influence of the variation of the fundamental constants on the predictions of BBN and extended our previous analysis~\cite{Coc07}.
Through our detailed modeling  of the cross-sections
we have shown that although the variation of the nucleon-nucleon potential
can greatly affect the triple-$\alpha$ process in stars, its effect on BBN and the production
of heavier elements such as CNO is minimal at best. At the temperatures, densities and 
timescales associated with BBN, the changes in the \aaag\ and \beac\ reaction rates
are not sufficient to produce more than C/H $\sim 10^{-21}$, 
and is therefore typically 6 orders of magnitude smaller than standard model abundances. 
This conclusion holds even
when including the possibility that $^8$Be can be bound.
  Using the variation of the fundamental constants provides a physically motivated and well-defined model to allow for stable \be8.

We have also extended previous analysis by including effects involving \he5 and \li5. 
This allowed us to revisit the constraints obtained in Ref.~\cite{Coc07} and in particular to show that the effect on the cross-sections remain small compared to the induced variation of $B_D$. This analysis demonstrates the robustness of our previous analysis and places the understanding of the effect of a variation of fundamental constants on BBN on a safer ground.

Our analysis can be compared with Ref.~\cite{Ber10} who reached the conclusion that such variations may increase the variation of \li7 and exacerbate the lithium problem. While formally correct, our results show that this required large variation of $\delta_{NN}$ is incompatible with the BBN constraints. Note also that Ref.~\cite{Che11} assumed an independent variation of the energies of the resonances while our work considers the variation of the energies of these resonance that arises from the same physical origin, so that their amplitudes are correlated.

Finally, we have extended our analysis to include the possibility of an extra relativistic degree of freedom.  Because of the different 
dependencies, Y$(\delta_{NN}, N_\nu)$ and D/H$(\delta_{NN}, N_\nu)$, the limits on 
$\delta_{NN}$ or $\delta h/h$ are not relaxed for $N_\nu = 4$. The only possibility 
to reconcile \li7 in this context is a variation of $\delta_{NN} \sim -0.01$, along with the post BBN
destruction of D/H.

\acknowledgements
 
The work of KA was supported in part by DOE grant
DE--FG02--94ER--40823 at the University of Minnesota. 
This work was also supported by the PICS CNRS/USA and the French ANR VACOUL.


\end{document}